# The Michigan Infrared Test Thermal ELT N-band (MITTEN) Cryostat

R. Bowens[*a], E. Viges[b], M. R. Meyer[a], D. Atkinson[a],
J. Monnier[a], M. Morgenstern[a], J. Leisenring[c] and W. Hoffmann[c]

[a]Department of Astronomy, University of Michigan, 1085 S. University, Ann Arbor, MI 48103;
[b]Space Physics Research Lab, University of Michigan, 2455 Hayward St., Ann Arbor, MI 48109;
[c]Steward Observatory, The University of Arizona, 955 N. Cherry Ave, Tucson, AZ 85719

## ABSTRACT

We introduce the Michigan Infrared Test Thermal ELT N-band (MITTEN) Cryostat, a new facility for testing infrared detectors with a focus on mid-infrared (MIR) wavelengths (8-13 microns). New generations of large format, deep well, fast readout MIR detectors are now becoming available to the astronomical community. As one example, Teledyne Imaging Sensors (TIS) has introduced a long-wave Mercury-Cadmium-Telluride (MCT) array, GeoSnap, with high quantum efficiency (> 65 %) and improved noise properties compared to previous generation Si:As blocked impurity band (BIB) detectors. GeoSnap promises improved sensitivities, and efficiencies, for future background-limited MIR instruments, in particular with future extremely large telescopes (ELTs). We describe our new test facility suitable for measuring characteristics of these detectors, such as read noise, dark current, linearity, gain, pixel operability, quantum efficiency, and point source imaging performance relative to a background scene, as well as multiple point sources of differing contrast. MITTEN has an internal light source, and soon an accompanying filter wheel and aperture plate, re-imaged onto the detector using an Offner relay. The baseline temperature of the cryostat interior is maintained < 40 K and the optical bench maintains a temperature of 16 K using a two-stage pulse-tube cryocooler package from Cryomech. No measurable background radiation from the cryostat interior has yet been detected.

**Keywords:** Infrared Detectors, Cryostat Design, Mid-infrared Instrumentation, Test facilities

## 1. INTRODUCTION

The performance of infrared detectors is key to the success of any infrared instrument designed for use at a telescope. Properties of such detectors are usually specified by the manufacturer in response to a request-for-quote (RFQ) from a customer, in this case a university research team, observatory, or national laboratory. It can be challenging for customers to verify these parameters in operation at a telescope. Further, the development of new detectors can benefit tremendously from analysis of data obtained in a controlled environment. MITTEN is designed to be flexible, enabling tests for a range of infrared detectors operating from 1-15 microns, at cryogenic temperatures between 20-100 K. Of particular interest to us are the new 3-13 micron MCT GeoSnap arrays from TIS (Jerrram & Beletic 2019)[1]. These devices benefited from investments in longwave MCT detectors for space-based applications (McMurtry et al. 2013 and references therein)[2] and are extremely promising for ground-based mid-infrared astronomy. At the moment these are the detectors of choice for future instrumentation on the next generation of extremely large telescopes (e.g. Brandl et al. 2018; Packham et al. 2018)[3,4]. Accurate knowledge of a detector's capabilities is key to working with the vendor to improve performance, instrument design, and formulating plans for operation at the telescope. In this contribution, we describe the design, implementation, and performance to date of the MITTEN cryostat. In section 2, we outline the science requirements that drove technical specifications to support design choices. In section 3, we provide an overview of the cryostat design (thermal mechanical, optical, and the light source/aperture/filter wheel assembly), and in section 4 we describe the performance to date of the cryostat, as well as future plans.

## 2. SCIENCE REQUIREMENTS INFORMING THE DESIGN

The MITTEN cryostat was specifically designed to enable testing of the GeoSnap longwave MCT detector from TIS (Jerram & Beletic 2019)[1]. Thus, the properties of the GeoSnap device drove the development of some of our requirements as well as the translation of those into technical specifications. However, we also envision its use in testing 1-5 micron MCT detectors such as the HAWAII-2RG detectors available from TIS. The University of Michigan GeoSnap detector, developed in collaboration with the University of Rochester, TIS, and the University of Arizona, has the following properties: a) 2048x2048 format read-out integrated circuit (ROIC) with 18 micron pixels of which one quadrant of 1024x1024 is photo-sensitive; b) maximum frame-rate of 85 Hz generating about 1 Watt of power during operation; c) operating temperature of approximately 30-50 K; d) read noise of approximately 150 e- in low gain mode; e) well depth of about 1.2 million e- (low gain); and f) quantum efficiency (QE) > 0.5 from 3-13 microns (un-thinned CdZnTe substrate without anti-reflection coating). Our system should be compatible with future devices that are operational over the full 2048x format, as well as exhibiting QE > 85 % (assuming thinned devices with anti-reflection coating).

---

[*] rpbowens@umich.edu; https://sites.lsa.umich.edu/feps/

### 2.1 Scientific Requirements

MITTEN should permit measurement of the following properties of GeoSnap detectors: a) dark current at operating temperatures between 30-50 K with a requirement of 10 % accuracy including uncertainties in the gain conversion of counts to electrons (goal 3 %).; b) gain (requirement of 10 % accuracy, goal of 3 % in electrons/ADU); c) read noise with a requirement of 10 % accuracy (goal of 3 %); d) linearity; e) pixel operability, and f) image quality of an unresolved point source on the detector, sampling the point spread function (PSF) full-width at half maximum (FWHM) with at least three pixels (requirement) at 10 microns with a goal of sampling the PSF with 10 pixels. Stretch goals for the system are to measure multiple point sources across the array, at contrasts > 100:1, as well as quantum efficiency to 15 % accuracy over > 25 % of the area of the array in at least three bands between 3.6-4.0, 4.8-5.0, and 10-12.5 microns.

### 2.2 Technical Specifications

Anticipating use of the mean variance method (Mortara & Fowler 1981)[5] to measure the gain at high signal to noise, we then specified the properties of a light source. Given the well-depth of the detector and the framerate, we benchmarked our technical specifications relative to a target source flux of half-full well (600,000 e-) within 0.0125 seconds. Given this signal level and expected shot noise (774 e-), we set other requirements. The thermal background of the cryostat (assuming uniform surface brightness over a solid angle of $2\pi$ steradians) should be low enough that it contributes less than 10 % to the noise to the measurements of dark current and light source flux on the detector. This led to the specification of the ambient temperature inside the cryostat to be less than 60 K. To minimize external leaks and stray light MITTEN does not have a window for light to enter from the outside: an internal light source is required. Given our interest in testing a range of detectors in the future, the cryostat should also be compatible with a wide range of external cable requirements.

## 3. CRYOSTAT DESIGN

Next, we outline the choices made to achieve our baseline design. We begin with the overall thermal mechanical design, followed by discussion of the optical design. We then discuss the design of the light source, aperture plate, and filter wheel assembly.

### 3.1 Thermal Mechanical

The cryostat was designed at the University of Michigan and built by Universal Cryogenics of Tucson, Arizona. MITTEN features an inner working volume that is 36.3 cm tall and 67 cm in diameter. The cryostat was designed such that the optical bench would reach temperatures < 20 K with the interior surfaces visible to the bench reaching < 60 K. There are no windows in the cryostat. The system includes two thermal shrouds. There are multiple O-ring sealed flanges on the outer shield that allow access to the interior for wiring. The interior optical bench features 1" spacing for ¼-20 tapped holes which can be configured in a variety of ways. Finally, there are four access plates for wire harnesses to accommodate diverse cabling requirements. The inner volume is approximately 0.128 meters cubed (128 liters). Cooling is achieved via a pulse-tube cryo-cooler as described below.

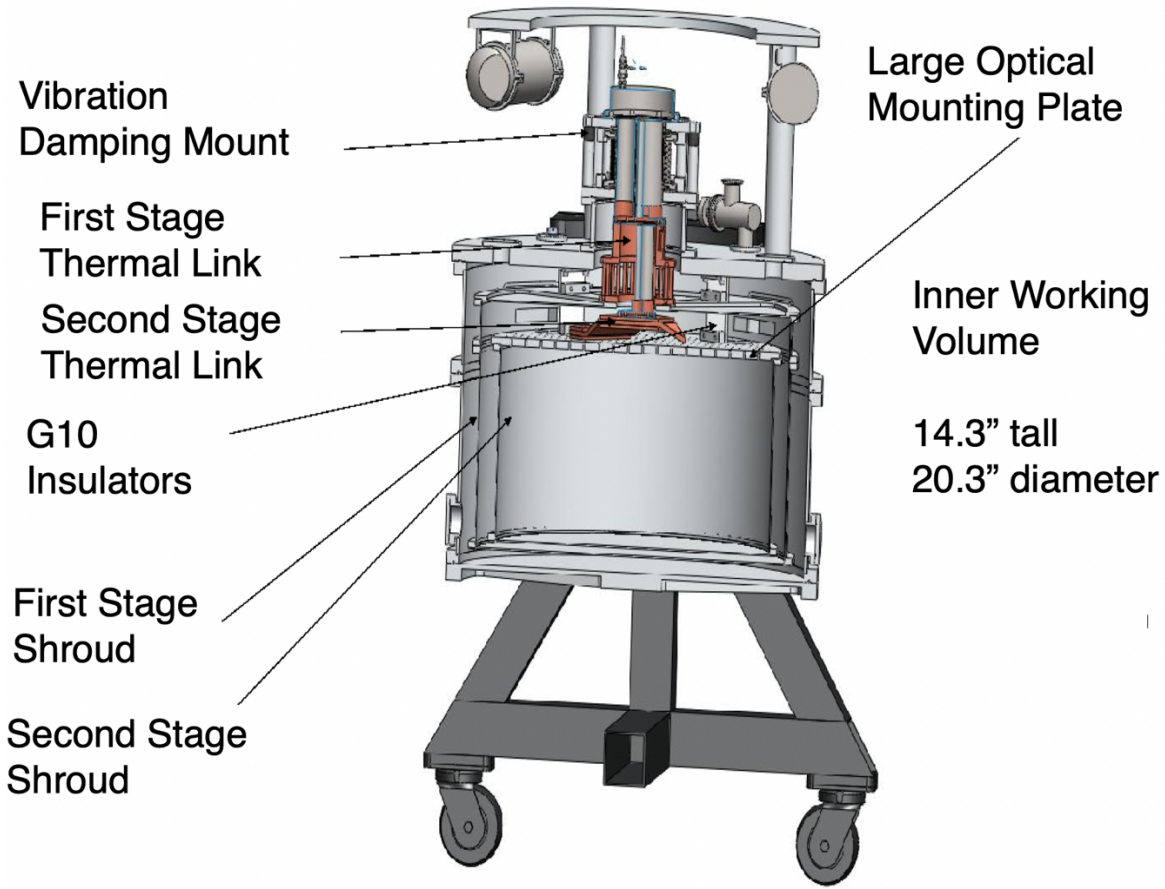

Figure 1. Cross sectional view of MITTEN as it rests on its cart during normal operation, with the pulse-tube protruding from the top of the cryostat for optimum performance.

Thermal properties were estimated using the SolidWorks software. With the built-in thermal properties of various materials, we explored thermal conductivity in the design. We ignored convection in these estimates since the vacuum should prevent any meaningful transfer. There are currently seven cryogenic temperature sensors installed throughout MITTEN. These feedback to either a Lakeshore Model 336 Temperature controller or a Lakeshore Model 224 Temperature Monitor which in turn control heaters on the light source and detector mount. Pressure and temperature data are stored electronically in a log file on a standard desktop computer. In the event of power outages, backups are in place to keep the Lakeshore devices operational. The cryocooler can be shut-off automatically and heating elements can be enabled to protect detectors from cooling too quickly or from reaching a lower limit in temperature. A battery can supply the 500-ohm heater with 20 volts (producing 0.8 watts) for 24 hours. Exact locations of sensors and heaters vary but baseline locations include the light source mounting block, light source, light source aperture slide, main optical plate, detector mounting block, and the stage one radiation shield.

The MITTEN cryostat uses an exterior shield, superinsulation (which protects the cold plate from radiative heating), an aluminum stage one shield, and an aluminum stage two shield in order to maintain critical temperatures (ambient, light source, detector). The first stage of the pulse tube cools the superinsulation and the stage 1 shield. The second stage of the pulse tube cools the stage 2 shield, allowing it to reach temperatures < 20 K. A variety of O-ring sealed flanges line the outside of the exterior shield, allowing access to the interior from different points. A vacuum valve can be connected to external pumps in order to pump down the interior. A rough pump is used in combination with a turbo pump to produce the interior vacuum. The cryostat uses a PT-810 PT-RM Cold Head which requires 12 W at 20 K or 72 W at 80 K. For the GeoSnap detector, a warm-to-cold cable of 22" length runs from a warm electronics box, through a potted feedthrough, to the interior of the cryostat.

The detector mount is a molybdenum block attached to 6061 aluminum. The thermal capacity is temperature dependent so with a 20 K base temperature, the molybdenum can reach +12.5 K to +21.9 K above the base (for 1 W or 2 W power in-flow, respectively). The detector generates one watt while the light source described below generates another watt. We can maintain the detector mount temperature to within +/- 0.001 K from 30-70 K. Care was taken so that detector cannot change temperature faster or slower than 1 K/minute. We also insure that the detector cannot drop below a minimum set temperature (currently 35 K) consistent with a validated thermal model for the GeoSnap detector mount from TIS: it can likely operate safely at lower temperatures though eventually the silicon-based ROIC will no longer function properly due to charge carrier freeze-out (cf. Rieke 2007)[6].

The cryostat sits on a cart where the "bottom" can be removed vertically with a small hand-operated crane. This permits easy access to the optical bench and working surfaces, while the cold head is upside down on the "top". During normal operation the cryostat is rotated such that the cold head is at the top, for optimum performance, and the optical bench is upside down (see Figure 1 shown during normal operation). In Figure 2 (a and b), the cryostat is shown open and "upside down" in order to access the optical bench.

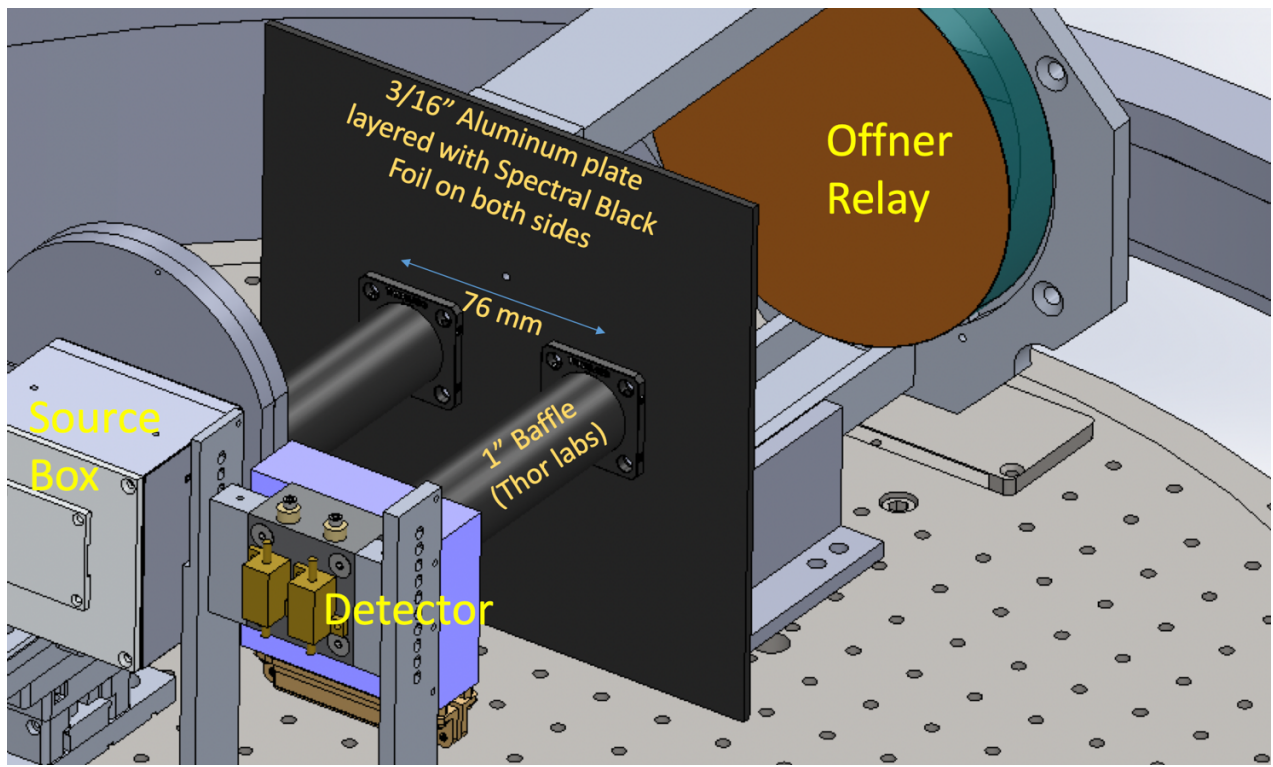

Figure 2a. The optical bench is shown with Offner relay (only 6" primary/tertiary visible) and detector mount, as well as light source and filter wheel assembly (round, in front of source box) with linear stages and aperture plate not visible. Baffles and shield tubes used to minimize scattered light are shown in black.

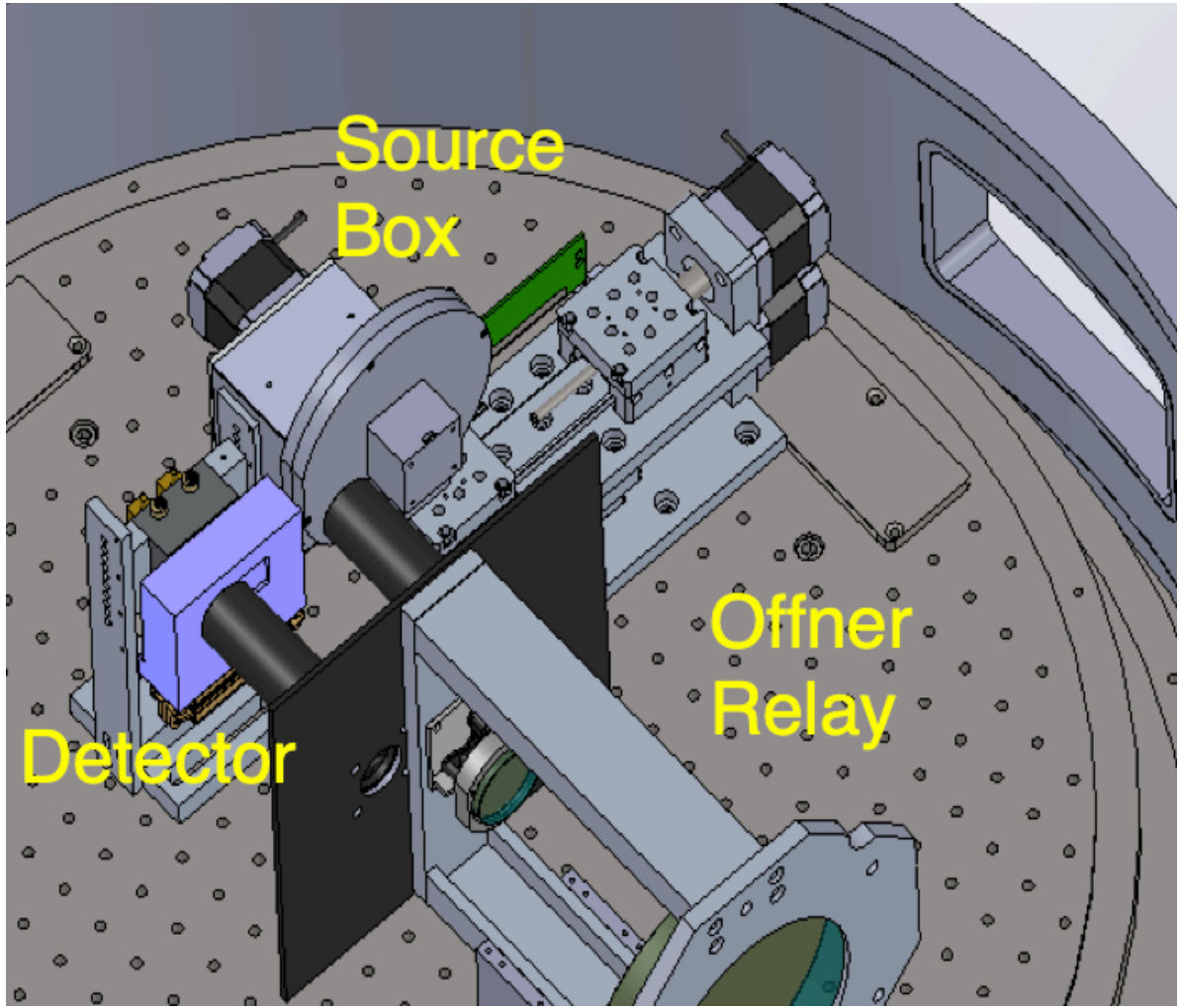

Figure 2b. Another view of the optical bench, this time with both mirrors of the Offner relay visible. The aperture slide in green and linear stages are also visible in this view.

### 3.2 Optics

In order to produce an image of an extended field (e.g. constant surface brightness light source, or any desirable scene, e.g. constellation of point sources) on the array, reimaging optics are required. This is accomplished via an Offner relay, first reflecting light from the source off a convex mirror, to a concave secondary mirror, and then back off the convex primary as shown in Figure 3. The Offner relay cancels out nearly all aberrations induced by the reflection off of the primary due to the inverse process on the third reflection. The relay has a 1:1 magnification. The design of the Offner relay requires that the radius of curvature of the secondary is half that of the primary (and opposite in sign).

Our science requirements dictated use of a primary greater than 3" diameter given our desired diffraction limit and light gathering power. An off-the-shelf set of primary and secondary mirrors were not available from commercial vendors. However, Graham & Treffers (2001)[7] found that one could use a convex lens with a gold coating on one side as the secondary in order to function as a cost-effective convex mirror. We were able to utilize an Offner relay constructed in this way, that had been designed for a previous infrared instrument at the University of Michigan. The primary is 6" and the distance between the two mirrors is 12". In Figure 3, the secondary has a stop defining an f/11 beam. This limits light to the detector by approximately 50% but keeps the beams separated given the limited area of the optical bench.

We examined the image quality of the spot diagrams generated in the Zeemax ray-tracing software. The spot diagrams are made from a 3 x 3 grid of points representing a 1 cm x 1 cm light source as an extended field approximately the size of the full 2048x GeoSnap array. The image quality of the f/11 beam over the field is roughly diffraction-limited (40" sampled across 4 pixels at full-width-half-maximum).

Finally, we note that hexagonal black light trap sheets (aluminum painted with a commercial black coating) are used as wall baffles to help minimize scattered light, as well as shield tubes for the rays as shown in Figure 2 (a and b).

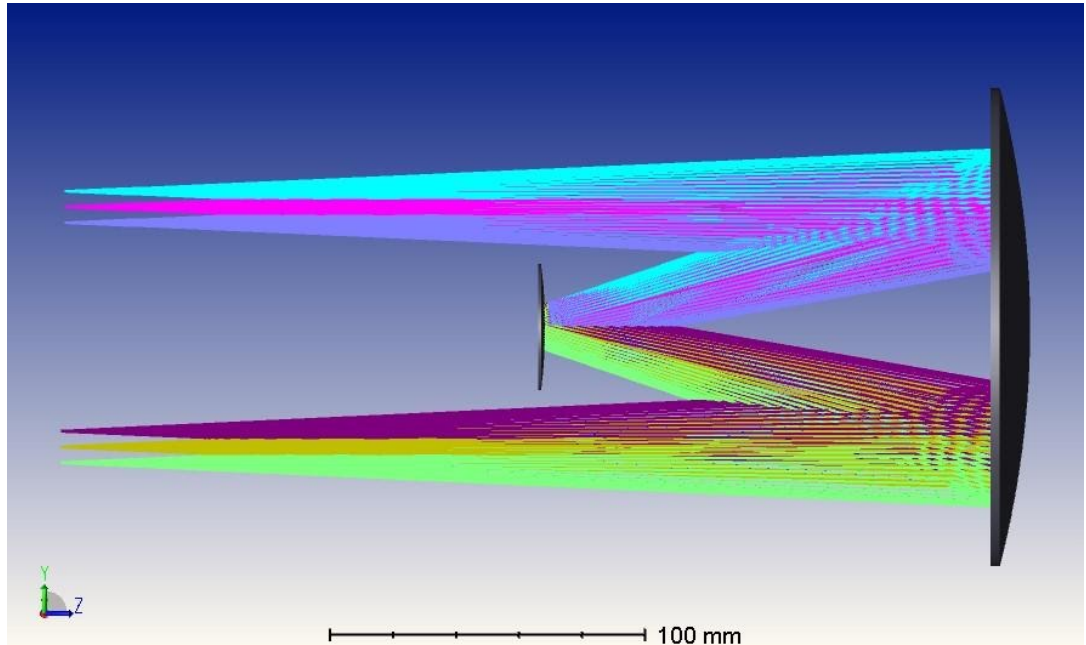

Figure 3. Zeemax ray tracing of the Offner relay with f/11 stop at the second mirror. Light source is at the top and detector is at the bottom in this projection.

### 3.3 Light Source, Aperture Plate, and Filter Wheel Assembly

In order to test detectors, a well-controlled source is needed. We considered use of an integrating sphere with an illumination pattern calibrated from moveable NIST-traceable single element photodiodes (Carter et al. 2006)[8]. However, the cost and schedule requirements led us to abandon this approach. Currently, 10 micron NIST-calibrated photodiodes are considered a research effort thus not yet routinely available.

We summarize the requirements of the light source as follows: a) produce an isothermal surface that provides a uniform blackbody radiation pattern (~1" x 1") to be imaged 1:1 onto the detector; b) capable of operating up to a max temperature of 400 K in order to emit sufficient radiation in the 3.6-4.0 micron band; c) use < 1 watt of power to reach 400 K in a < 20 K environment; d) cool fast enough from 400 K to 40 K as to be useful within 48 hours; e) allow for switching between filters by use of a filter wheel; and f) capable of controlling the temperature of a moveable aperture plate (with point sources produce by holes < 8 microns in diameter) such that the bright point source will appear upon a uniform background via the pinholes. This corresponds to a temperature of approximately 140 K for the slide, with the main source reaching temperatures > 200 K as seen through the pinholes.

A pure electrolytic copper (99.99% Cu alloy 101) is used for the main source since it has a high thermal conductivity (~400 W/mK) yet is inexpensive compared to alternatives like silver or diamond. The high thermal conductivity ensures that the source maintains a minimal temperature gradient. The copper is 8mm thick which balances the minimal temperature gradient (which decreases with thickness) with the cooling time. Finally, the front face of the copper is coated in ultra-black nano particles (from a commercial vendor) to achieve high emissivity.

Thin G10 insulator is used to suspend the source within the aluminum source box. The exact shape of the G10 suspension can be modified with different cuts in order to increase or decrease the thermal isolation of the source. Figure 4 shows the basic design of the source enclosure, copper source block, and the heating element.

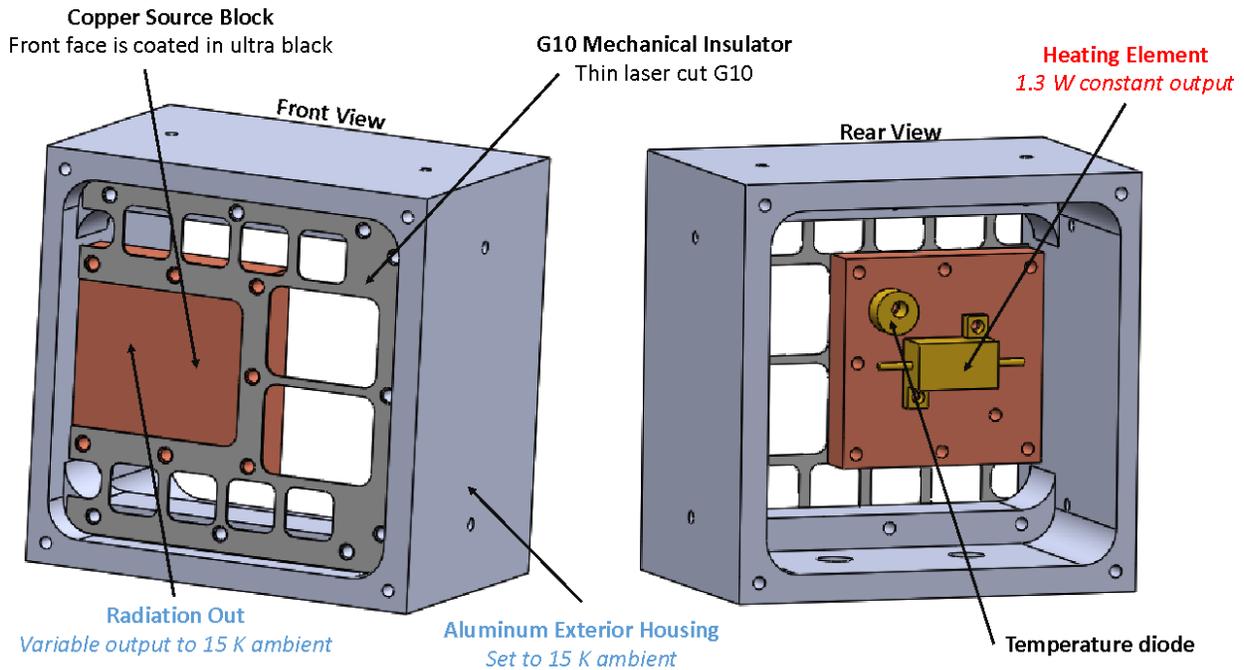

Figure 4.  Temperature controlled light source.

We also used simulations in Solidworks to study a static thermal model of the source. We included the impact of conduction and radiation. Since the cryostat is a vacuum, we did not include convection. The maximum operating temperature of the source (400 K), should be worst case for uniformity in the copper source block since thermal gradients between the source and source housing are most severe.  A model from our simulation is visible in the following figure.  Across the source, the temperature varied by a max of 0.05 K. For a wavelength band of 3.6 to 4.0 micrometers with a source emissivity of 0.9, we find the percent change in flux at 400 K is only 0.044%.

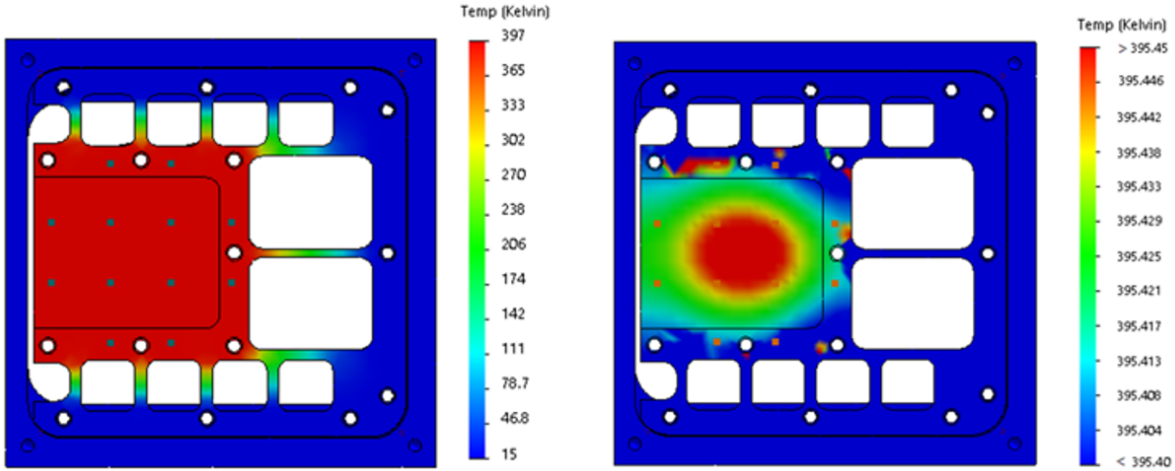

Figure 5. Thermal modeling of source in a 15 K environment shown on the left and thermal gradients within source on the right.

The source box features a sliding pinhole. The pinhole features surface mount resistors in order to allow for heating of the slide area around the pinhole (which is necessary for producing the background). There is also a temperature sensor attached so that the pinhole's temperature can be monitored. A filter wheel is located in front of the pinhole mechanism. The wheel can rotate through six positions which currently hold the following filters: 3.6-4.0 microns, 4.93-5.13 microns, 10.1 to 11.3 microns, 7.75 to 9.98 microns, an open position, and a sealed position. Resistors are used to track the position of the wheel. Finally, the whole source box sits on a series of three linear stages that allow one to adjust the focus towards or away from the Offner relay (first large mirror), the pinhole location (shifting from broad illumination to various pinhole illuminations), or the source as projected onto the detector (left or right). The latter motion will enable us to sense the exact same area of the light source on two different parts of the detector. The light source, aperture slide, filter wheel as associated cryogenic stepper motor driven stages are shown in Figure 2 (a (center left) and b (center)).

## 4. PERFORMANCE TO DATE AND FUTURE PLANS

The main cryostat hardware was delivered in August, 2019. Following assembly and installation in our lab, the cryostat was pumped down to $<10^{-4}$ torr. With vacuum hold verified, and temperate monitors enabled, the cryostat was cooled down, again without incident. The University of Michigan GeoSnap detector was installed in early 2020 and has been under testing since that time. Results from these tests will be reported elsewhere (Leisenring et al.; Atkinson et al. in preparation). The cryostat takes approximately 24 hours to cool down or heat up to room temperature relative to an ambient temperature of 16 K for the optical bench. This temperature is reached with 1 watts of power input required by the detector. At this equilibrium operating temperature, the interior of the cryostat is at 40 K (sides and top measured). To date, we have not been able to measure any ambient thermal background within the detector, although our dark current measurements appear to be limited by amplifier glow originating from the detector and ROIC.

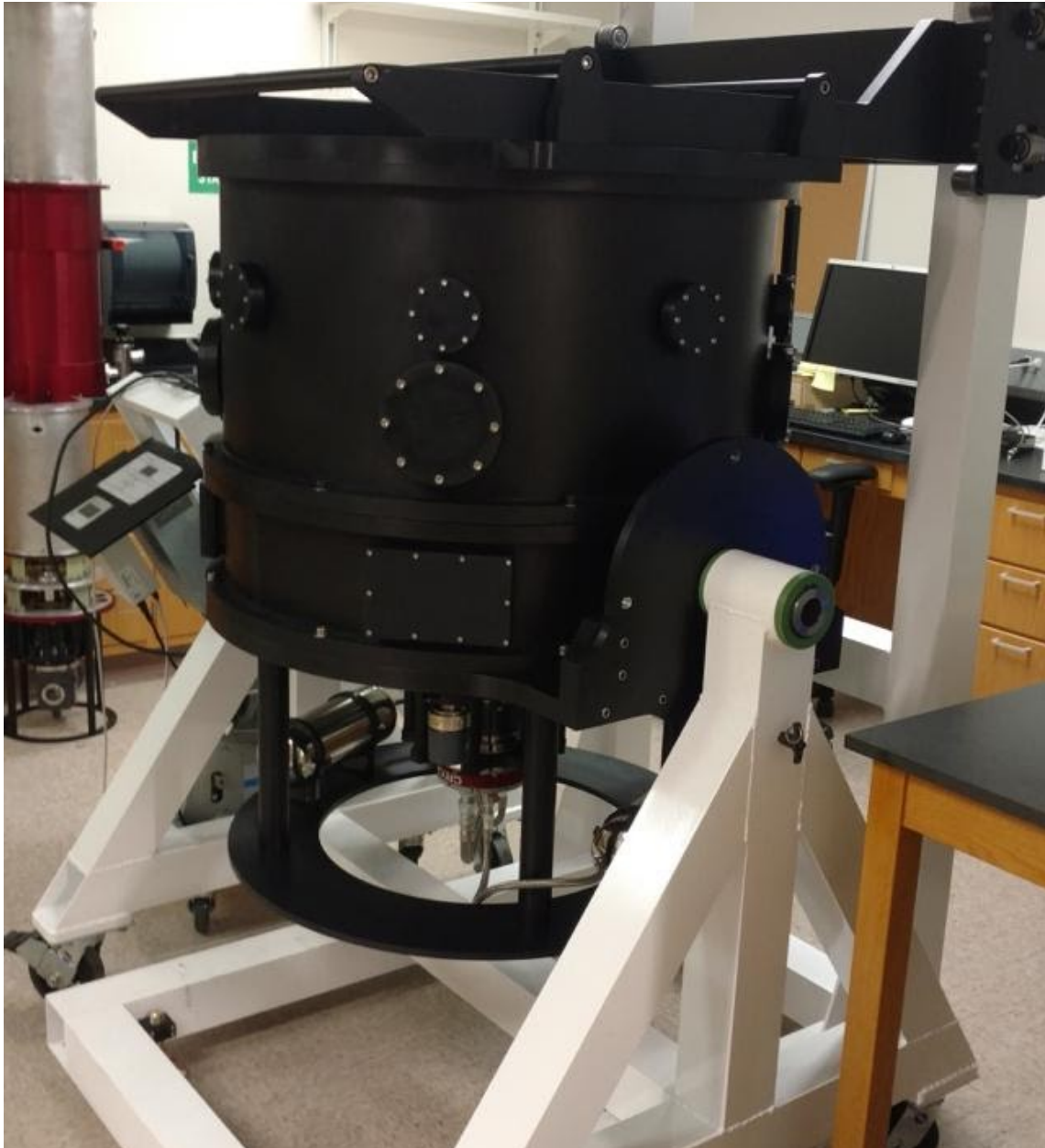

Figure 6. External view of the MITTEN cryostat "upside down" with the pulse-tube underneath, ready for removing the "bottom" with the crane, enabling work on the optical bench.

We currently have installed a temporary light source, consisting of a metal screen roughly the size of the array, on a post placing it at the height of the detector, one inch apart, with a heater attached. Thermal models indicate temperatures of < 50 K to > 400 K. We have calibrated the voltage supplied to the heater needed to produce half-well images for a variety of framerates. Due to the non-uniform illumination of the array, center to edge flux differences of more than x2 are observed.

The main components of the optical bench including the light source, aperture plate stage, filter wheel, Offner relay, and baffling have yet to be installed in the cryostat. This is in part due to urgent testing done in collaboration with TIS and other stakeholders in the GeoSnap detector development for ground-based astronomy. To date, we have been able to measure dark current, gain, read noise, linearity, and pixel operability with MITTEN and the temporary light source described above. All components of the optical bench are now in hand. The assembly of the motion stages, blackbody source, filter wheel, and aperture plate with pinhole point-sources is completed. Integration of the optical components and system control software into MITTEN is ongoing (Figure 6). These components will be installed early in 2021 to enable further tests of the GeoSnap detector such as quantum efficiency and response to point sources relative to thermal background, as well as multiple point sources with contrasts of > 30:1 at < 10 $\lambda/D$.

Acknowledgements

This project was made possible through the support of a grant from the Templeton World Charity Foundation, Inc. The opinions expressed in this publication are those of the authors and do not necessarily reflect the views of the Templeton World Charity Foundation, Inc. We are also thankful for support from the University of Michigan that enabled the founding of the Infrared Imaging and Spectroscopy (IRIS) Lab in the Formation and Evolution of Planetary Systems (FEPS) research group. We are also grateful for the collaboration of our colleagues at Universal Cryogenics, the University of Michigan Department of Physics, in particular Paul Thurmond and his team in Randall Laboratory, the Space Physics Research Lab, and the Plasmadynamic and Electric Propulsion Lab, as well as at Steward Observatory, The University of Arizona including Manny Montoya and Dennis Hart. Finally, we acknowledge encouragement and support from Teledyne Imaging Sensors, in particular Vincent Douence, John Auyeung, and Jim Beletic.